\documentclass[twocolumn,prb,showpacs,color,superscriptaddress,epsfig]{revtex4}
\usepackage{graphicx}
\usepackage{amsfonts}
\usepackage{amsmath}
\usepackage{amssymb}
\usepackage{epsfig}
\usepackage{color}
\begin{document}
\title{Cross-relaxation and phonon bottleneck effects on magnetization dynamics in LiYF$_4$:Ho$^{3+}$}

\author{S. Bertaina}
\affiliation{Laboratoire de Magn\'etisme Louis N\'eel, CNRS, BP 166,
38042 Grenoble CEDEX-09,France}\affiliation{ Ecole Nationale
Sup\'erieure de Physique de Grenoble,38402, St-Martin
d'H\`eres,France}

\author{B. Barbara}
\affiliation{Laboratoire de Magn\'etisme Louis N\'eel, CNRS, BP 166,
38042 Grenoble CEDEX-09,France}

\author{R. Giraud}
\affiliation{Laboratoire de Magn\'etisme Louis N\'eel, CNRS, BP 166,
38042 Grenoble CEDEX-09,France}\affiliation{LPN, CNRS,
91460-Marcoussis, France}

\author{B.Z. Malkin}
\affiliation{Kazan State University, Kazan 420008, Russian
Federation}

\author{M.V.Vanuynin}
\affiliation{Kazan State University, Kazan 420008, Russian
Federation}

\author{A.I. Pominov}
\affiliation{Kazan State University, Kazan 420008, Russian
Federation}

\author{A.L. Stolov}
\affiliation{Kazan State University, Kazan 420008, Russian
Federation}

\author{A.M. Tkachuk}
\affiliation{S.I. Vavilov State Optical Institute, St.Petersburg
199034, Russian Federation}

\date{\today}

\begin{abstract}
Frequency and dc magnetic field dependences of dynamic
susceptibility in diluted paramagnets LiYF$_4$:Ho$^{3+}$ have been
measured at liquid helium temperatures in the ac and dc magnetic
fields parallel to the symmetry axis of a tetragonal crystal
lattice. Experimental data are analyzed in the framework of
microscopic theory of relaxation rates in the manifold of 24
electron-nuclear sublevels of the lowest non-Kramers doublet and the
first excited singlet in the Ho$^{3+}$ ground multiplet $^5I_8$
split by the crystal field of S$_4$ symmetry. The one-phonon
transition probabilities were computed using electron-phonon
coupling constants calculated in the framework of exchange charge
model and were checked by optical piezospectroscopic measurements.
The specific features observed in field dependences of the in- and
out-of-phase susceptibilities (humps and dips, respectively) at the
crossings (anti-crossings) of the electron-nuclear sublevels are
well reproduced by simulations when the phonon bottleneck effect and
the cross-spin relaxation are taken into account.

\end{abstract}


\maketitle

\section{Introduction}

In 1968 Hellwege et al. \cite{Hellweg1968} found an unusual non
monotonous behavior of the in-phase magnetic susceptibility in
parallel dc and ac magnetic fields with narrow maxima at the dc
field values corresponding to the crossings of electron-nuclear
sublevels of the ground-state of diluted Ho$^{3+}$ ions in LaCl$_3$
at the temperature $\sim$1 K and frequencies of the ac-field in the
range $10^2-5.10^3$ Hz. Recently, similar data were independently
obtained on the dynamic susceptibility of the diluted paramagnet
LiYF$_4$:Ho$^{3+}$ (0.1\%) in the region of energy level crossings
in Ref. [\onlinecite{Giraud2003}]. This latter study of the
anisotropic rare-earth spin dynamics in the classical regime was an
extension of sub-Kelvin magnetization measurements, previously used
to reveal the role of quantum fluctuations of atomic magnets in the
diluted LiYF$_4$:Ho$^{3+}$ system \cite{Giraud2001}, similarly to
the phenomenon of resonant tunneling of single molecule magnets in
presence of a large uniaxial anisotropy \cite{Thomas1996}. In both
[\onlinecite{Hellweg1968}] and [\onlinecite{Giraud2003}] the
observed peculiarities of the susceptibility dynamics, measured at
relatively high temperature, relate to predictions of N. Bloembergen
and co-workers in their classical study of the cross-relaxation in
spin systems \cite{Bloembergen1959}: "\emph{...the susceptibility is
usually plotted at constant frequency versus applied dc field. There
may easily occur a maximum in this plot, because for certain values
of the external field some pairs of levels ... may become nearly
equidistant}". The purpose of the present paper is to investigate
the microscopic origin of the susceptibility in diluted
LiYF$_4$:Ho$^{3+}$ at liquid helium temperatures, taking into
account the different effects of crystal-field, electron-phonon and
hyperfine interactions, as well as cross-relaxation processes. The
extension of this many-body classical dynamics to the
low-temperature quantum case was previously adressed in
Refs.[\onlinecite{Giraud2003,Giraud2001}]. Nevertheless, many-spin
quantum fluctuations in highly anisotropic systems, such as
co-tunneling processes, have not been discussed on microscopic
grounds up to now. Therefore, to go further in the analysis of
dynamical magnetic hysteresis loops measured at sub-kelvin
temperatures in the strongly out-of-equilibrium quantum regime
\cite{Giraud2001,barbara2004}, a clear understanding of the
classical spin dynamics in LiYF$_4$:Ho$^{3+}$ is required. The
quasi-resonant absorption of radiofrequency (5-10 MHz) ultrasound
and ac magnetic field energy at the crossing points in the
CaWO$_4$:Ho$^{3+}$ crystal was observed in
Ref.[\onlinecite{Grinberg1982}], and some specific peculiarities of
$^{19}$F nuclear relaxation rates at the crossing points in
LiYF$_4$:Ho$^{3+}$ were found in Ref.[\onlinecite{Graf2006}]. All
these experimental findings give evidence for essential variations
of the relaxation rates in the electron-nuclear subsystem, by orders
of magnitude, within the vicinity of energy level crossing points.
In the present work, new measurements of frequency, temperature and
dc magnetic field dependences of the dynamic susceptibility in
LiYF$_4$ single crystals containing different concentrations of
Ho$^{3+}$ ions have been carried out. The experimental data are
analyzed in the framework of the microscopic theory of the
electron-phonon interaction in a gapped titled system. The existing
data on spectral properties of Ho$^{3+}$ ions in LiYF$_4$
\cite{shakurov2005} and additional piezospectroscopic studies
described below allowed to obtain reliable values of electron-phonon
coupling constants and rigorous estimates of relaxation rates in the
manifold of lower electron-nuclear sublevels of the ground
multiplet. Simulations based on the calculated values of kinetic
parameters revealed remarkable differences between the computed and
measured dependences of the dynamic susceptibility on the external
parameters (frequency, temperature, strength of the constant
magnetic field). We found it necessary to derive a more thorough
theoretical approach accounting for the phonon bottleneck effect and
the cross-relaxation. A self-consistent description of all specific
features of the measured susceptibilities has been achieved using
parameters of the phonon and magnetic subsystems which have physical
meaning. The paper is arranged as follows: in the first section we
derive a general expression of the dynamic susceptibility, then we
describe the known spectral properties of impurity Ho$^{3+}$ ions in
LiYF$_4$ and calculate electron-phonon coupling constants which are
used to compute the relaxation matrix in the space of lower 24
electron-nuclear sublevels of the ground multiplet in the external
magnetic field. The next section contains results of experimental
studies, and in the last section we compare measured and simulated
in- and out-of-phase susceptibilities.

\section{Theory of dynamic magnetic susceptibility of a diluted paramagnet}

The dynamic susceptibility $\mathbf{\chi}(\omega)$ of a single
paramagnetic ion coupled to a phonon bath in the external magnetic
field $\mathbf{B}=\mathbf{B}_0+\mathbf{B}_1(\omega,t)$ $,
(\mathbf{B}_1(\omega,t)= \mathbf{B}^0_1 \exp(-i\omega t)$, is
determined by the following expression:

\begin{equation}\label{eq:1}
    \chi_{\alpha\beta}(\omega)=\frac   {Tr(\rho(t)\Delta m_\alpha)}
    {B_{1\beta}}.
\end{equation}

Here  $\Delta \mathbf{m}=\mathbf{m}-\langle \mathbf{m}\rangle_0$;
$\mathbf{m}$ is the operator of the ion magnetic moment, and
$\langle...\rangle_0$ defines an average value, corresponding to the
equilibrium single ion density matrix $\rho_0(H_0)=
\mathrm{e}^{-H_0/kT}/Tr(\mathrm{e}^{-H_0/kT})$ in absence of the
time dependent field. Time evolution of the density matrix $\rho(t)$
defined in the space of eigenfunctions $|k\rangle$ corresponding to
eigenvalues $E_k$ of the Hamiltonian $H_0$ of the unperturbed
electronic (or electron-nuclear) system can be described by the
master equation for diagonal elements $\rho_{nn}=\rho_n$ and by
simple exponential decay of non-diagonal matrix elements (validity
of this "secular" approximation was discussed, in particular, in
Refs. [\onlinecite{Leuer10,32Poh}]):

\begin{equation}\label{eq:2}
\frac{\partial \rho_n}{\partial t}=\sum_k W_{nk}\rho_k
\end{equation}

\begin{equation}\label{eq:3}
\frac{\partial \rho_{nk}}{\partial
t}=-(\gamma_{nk}+i\omega_{nk})\rho_{nk}+\frac{i}{\hbar}[\mathbf{mB}_1(\omega,t),\rho]_{nk}
\end{equation}

\begin{equation*}
\quad (n \neq k, \omega_{nk}=(E_n-E_k)/\hbar)
\end{equation*}

When working with eq.\eqref{eq:2} and \eqref{eq:3} we implicitly
suppose that the equilibrium state is established in the bath much
faster than in the electronic subsystem. Besides, we have omitted in
the r.h.s. of eq.\eqref{eq:2} the terms $-2\mathrm{Im}\sum_{k\neq
n}\mathbf{m}\mathbf{B}_1(\omega,t)\rho_{kn}/\hbar$,which do not
contribute to the linear response on the weak alternating field
$\mathbf{B}_1(\omega,t)$. The off-diagonal elements of the
relaxation matrix $W_{nk} =W_{k\rightarrow n}$ are the transition
probabilities induced by the electron-phonon interaction. For the
one-phonon transitions,$W_{m \rightarrow k}
=w_{mk}[n(\omega_{mk})+1]$  if the frequency of the transition
$\omega_{mk}>0$ and $W_{m \rightarrow k} =w_{mk}n(\omega_{km})$ if
$\omega_{mk}<0$ where $w_{mk}$ is the probability of the spontaneous
transition, and $n(\omega_{mk})$is the phonon occupation number. The
diagonal element $W_{nn}=-\sum_k W_{kn}$ determines the lifetime of
the corresponding state $n$, and the coherence decay rate equals

\begin{equation}\label{eq:5}
    \gamma_{nk}=-\frac{1}{2}(W_{nn}+W_{kk})+\Gamma_{nk}
\end{equation}

where the first term is caused by finite lifetimes, and
$\Gamma_{nk}$ determines all additional contributions to the
homogeneous broadening of the $n \rightarrow k$ transition.

Considering the energy of interaction with the time-dependent field
$-\mathbf{mB}_1(\omega,t)$ as a perturbation, we solve equations of
motion in the linear approximation:
\begin{equation}\label{eq:4}
    \rho_n(t)=\rho_{0n}+\Delta\boldsymbol{\rho}_n(\omega)\mathbf{B}_1\exp(-i\omega t)
\end{equation}
\\where $\Delta\boldsymbol{\rho}_n(0)=\rho_{0n}(\mathbf{m}_{nn}-\langle \mathbf{m}\rangle_0)/kT$

Postulating that the interaction with the phonon bath tends to
establish an equilibrium state at the instant value of the magnetic
field,$\sum_k W_{nk}\rho_{0k}(H_0-\mathbf{mB}_1)=0$ , we obtain from
eqs. \eqref{eq:2}\eqref{eq:3} and \eqref{eq:5}

\begin{equation}\label{eq:6}
    \Delta\rho_{\beta,n}(\omega)=\sum_{kp}(i\omega 1 +
    \mathbf{W})^{-1}_{nk}W_{kp}\Delta m_{\beta,pp}\rho_{0p}/kT
\end{equation}

\begin{equation}\label{eq:7}
    \Delta\rho_{\beta,nk}(\omega)=\frac{m_{\beta,nk}(\rho_{0k}-\rho_{0n})}{\hbar(\omega_{nk}-\omega-i\gamma_{nk})}
\end{equation}

The dynamic susceptibility \eqref{eq:1} takes the form

\begin{eqnarray}\label{eq:8}
\chi_{\alpha\beta}(\omega)&=& \chi_{\alpha\beta}^0
-\{i\omega\sum_{nk}\Delta m_{\alpha,nn}(i\omega 1 +
\mathbf{W})^{-1}_{nk}\nonumber\\
&&\times\Delta
m_{\beta,kk}\rho_{0k}/kT\} \nonumber\\
&&+\sum_{n,k\neq
n}m_{\alpha,nk}m_{\beta,kn}(\rho_{0k}-\rho_{0n})\nonumber\\
&&\times\left(
\frac{1}{\hbar(\omega_{nk}-\omega-i\gamma_{nk})}-\frac{1}{\hbar\omega_{nk}}\right)
\end{eqnarray}

where the well known expression for the static susceptibility

\begin{eqnarray}\label{eq:9}
    \chi^0_{\alpha\beta}&=&\sum_n \Delta m_{\alpha,nn}\Delta
    m_{\beta,nn}\rho_{0n}/kT\nonumber\\
    &&+\sum_{n,k\neq n}\frac{m_{\alpha,nk}m_{\beta,kn}}{\hbar\omega_{nk}}(\rho_{0k}-\rho_{0n})
\end{eqnarray}
has been used. The first line in eq. \eqref{eq:8} coincides with the
expression presented in Ref. \onlinecite{Santini2005}.

In the case of a finite concentration of paramagnetic ions, we have
to take into account interactions between the ions (dipole-dipole,
dimer or trimer exchange, virtual phonon exchange) and the finite
rate of a heat flow from the phonon reservoir to the helium bath
(the phonon bottleneck effect). These interactions between
paramagnetic ions induce energy exchange (cross-relaxation), and the
master equation \eqref{eq:2} contains additional nonlinear terms
\cite{Bloembergen1959}:

\begin{eqnarray}\label{eq:10}
    \dot{\rho}_n&=&\sum_m W_{nm}\rho_m+\sum_{mpl}
    [(W_{np,lm}^{CR}\rho_p\rho_m-W_{pn,ml}^{CR}\rho_l\rho_n)\nonumber\\
    &&-(W_{np,lm}^{CR}\rho_p\rho_m-W_{pn,ml}^{CR}\rho_l\rho_n)_{ad}]
\end{eqnarray}
Here $W_{np,lm}^{CR}$ is the probability of the simultaneous
transitions $m\rightarrow l$ of one ion, and $p\rightarrow n$ of
another ion. According to the Fermi Golden rule, this transition
probability can be written as follows \cite{Bloembergen1959,Al1972}:

\begin{equation}\label{eq:11}
    W_{np,lm}^{CR}=\frac{2\pi}{\hbar}\left\langle\left|<n,l|H_{12}|p,m>\right|\right\rangle_{Av}\delta(\omega_{pn}-\omega_{lm})
\end{equation}
where $H_{12}$is the Hamiltonian of interaction between the ions,
and $\langle\rangle_{Av}$ means a configuration averaging over the
distribution of the paramagnetic ions in the crystal lattice. In the
particular case of dipole-dipole interactions between rare-earth
magnetic moments $\mathbf{m}=g_J\mu_B\mathbf{J}$( $\mu_B$ is the
Bohr magneton, $\mathbf{J}$ is the total angular momentum, $g_J$ is
the Lande factor),the Hamiltonian $H_{12}$ has the form
\begin{equation}\label{eq:12}
    H_{12}=\frac{(g_J\mu_B)^2}{R^3}[\mathbf{J}_1\mathbf{J}_2-3(\mathbf{J}_1\mathbf{R})(\mathbf{J}_2\mathbf{R})/R^2]
\end{equation}
where $\mathbf{R}$ is the vector connecting the ions. The last term
at the right-hand side of eq. \eqref{eq:10}, (...)$_{ad}$, provides
an asymptotical approach to the equilibrium distribution of
populations in the electronic system isolated from the phonon bath
(in adiabatic conditions) due to the cross-relaxation processes
\cite{Bloembergen1959}.

Taking into account the distribution of the single ion energies, we
can write the transition probability \eqref{eq:11} in the following
form:

\begin{eqnarray}\label{eq:13}
    W_{np,lm}^{CR}&=&\delta^2\sum_{\alpha\beta\gamma\delta}g_{\alpha\beta\gamma\delta}^{CR}
    (\omega_{pn}-\omega_{lm})k_{\alpha\beta\gamma\delta}\times\nonumber\\
    &&(<n|J_{1\alpha}|p><l|J_{2\beta}|m> \times \nonumber\\
    &&<p|J_{1\gamma}|n><m|J_{2\alpha}|l>+c.c.)
\end{eqnarray}
where $g_{\alpha\beta\gamma\delta}^{CR} (\omega)$ is the
cross-relaxation line shape function, $\delta$ is the average energy
of the interaction in frequency units (we can consider
$\delta=(g_J\mu_B)^2/\hbar R_0^3$ with $R_0$ equal to a lattice
constant as a scaling factor), and $k_{\alpha\beta\gamma\delta}$ are
the average values of the corresponding dimensionless lattice
factors. The Hamiltonian \eqref{eq:12} contains only symmetrical
products of the components of the angular moments, {$J_{1\alpha}
J_{2\beta}$}, and in a general case there are 21 independent
parameters $k_{\alpha\beta\gamma\delta}$. However, this number can
be essentially diminished due to symmetry properties of a concrete
lattice (similarly to the elastic compliance tensor). In particular,
in a case of S$_4$ local symmetry of paramagnetic ions, we obtain
\begin{eqnarray}\label{eq:14}
    &&W_{np,lm}^{CR}\delta^{-2}=\\
    &&g_{33}^{CR}k_{33}|J_{1z}^{np}J_{2z}^{lm}|^2\nonumber\\
    +&&g_{11}^{CR}k_{11}(|J_{1+}^{np}J_{2-}^{lm}+J_{1-}^{np}J_{2+}^{lm}|^2)\nonumber\\
    +&&g_{66}^{CR}k_{66}(|J_{1+}^{np}J_{2+}^{lm}-J_{1-}^{np}J_{2-}^{lm}|^2)\nonumber\\
    +&&g_{12}^{CR}k_{12}(|J_{1+}^{np}J_{2+}^{lm}+J_{1-}^{np}J_{2-}^{lm}|^2)\nonumber\\
    +&&g_{44}^{CR}k_{44}[(|(J_{1+}^{np}+J_{1-}^{np})J_{2z}^{lm}+J_{1z}^{np}(J_{2+}^{lm}+J_{2-}^{lm})|^2\nonumber\\
    +&&|(J_{1+}^{np}-J_{1-}^{np})J_{2z}^{lm}+J_{1z}^{np}(J_{2+}^{lm}-J_{2-}^{lm})|^2]\nonumber\\
    +&&g_{13}^{CR}k_{13}[J_{1z}^{np}J_{2z}^{lm}(J_{1+}^{pn}J_{2-}^{ml}+J_{1-}^{pn}J_{2+}^{ml})+c.c.]\nonumber\\
    +&&g_{16}^{CR}k_{16}[(J_{1+}^{np}J_{2+}^{lm}-J_{1-}^{np}J_{2-}^{lm})(J_{1+}^{pn}J_{2+}^{ml}+J_{1-}^{pn}J_{2-}^{ml})+c.c.]\nonumber
\end{eqnarray}
with $J^{ab}=<a|J|b>$. Because a similar expression may be obtained
in a case of long range interactions between the paramagnetic ions
through the field of elastic lattice deformations
($H_{12}^d=\sum_{pknm}A_{pk}^{nm}C_k^p(1)C_m^n(2)$ where $C_p^k(i)$
is the spherical tensor operating in the space of states of the ion
$i$, and $A_{pk}^{nm}$ are the coupling constants\cite{Baker1971})we
consider seven factors $k_{ab}$ introduced in \eqref{eq:14} as the
phenomenological parameters.

If the system is not far from equilibrium, equations \eqref{eq:10}
can be linearized. We suppose that the adiabatic density matrix is
characterized by a single parameter, the adiabatic temperature. The
difference

\begin{equation*}
    T^{-1}-T^{-1}_{ad}=\frac{\sum_nkE_n\Delta \boldsymbol{\rho}_n\mathbf{B}_1-T^{-1}\langle H_0\Delta \mathbf{m}\mathbf{B}_1\rangle_0}
    {\langle H_0^2\rangle_0-\langle H_0\rangle_0^2}
\end{equation*}
between the lattice and adiabatic temperatures is determined from
the condition that the average values of the electronic energy
$H_0-\mathbf{mB}_1$ obtained with the adiabatic density matrix and
with $\rho(t)$ are equal. The master equation in the form
\eqref{eq:2} and the expression \eqref{eq:8} for the dynamic
susceptibility remain valid but with the additional contributions to
the relaxation matrix

\begin{eqnarray}\label{eq:15}
    W_{nm}&=&W_{m\rightarrow n}\\
    &&+\sum_k W_{nk}^{CR}\left[\delta_{km}-\rho_{0k}\frac{(E_k-\langle H_0\rangle_0)(E_m-\langle H_0\rangle_0)}
    {\langle H_0^2\rangle_0-\langle H_0\rangle_0^2}\right]\nonumber
\end{eqnarray}
where

\begin{equation}\label{eq:16}
    W_{nm}^{CR}=\sum_{lp}(W_{np,lm}^{CR}\rho_{0p}+W_{nm,lp}^{CR}\rho_{0p}-W_{pn,lm}^{CR}\rho_{0n}),
    (n\neq m)
\end{equation}
Let us now take into account the finite relaxation rate of the
phonon subsystem \cite{VV1941} considering equations of motion for
the phonon occupation numbers along with the \eqref{eq:2}):
\begin{eqnarray}\label{eq:17}
\frac{dn(\omega_{mk})}{dt}&=&\frac{1}{\tau_{ph}(\omega_{mk})}[n(\omega_{mk})-n_0(\omega_{mk})]\\
&& +\frac{w_{mk}N}{P_{mk}\Delta
\omega_{mk}}(\rho_m[n(\omega_{mk})+1]-\rho_kn(\omega_{mk}))\nonumber
\end{eqnarray}
here $n_0(\omega_{mk})=[\exp(\hbar \omega_{mk}/kT)-1]^{-1}$ is the
equilibrium occupation number of phonons at the resonance frequency
$\omega_{mk}>0$ in the band with the width $\Delta\omega_{mk}$,
$\tau_{ph}$ is the phonon lifetime, $N$ is the number of
paramagnetic ions per unit volume, $P_{mk}$ is the density of states
of resonant phonons (for the low-frequency acoustic phonons $P_{mk}=
3\omega_{mk}^2/2\pi^2v^3$, where $v$ is the average sound velocity).
The stationary solution of the linearized coupled equations
\eqref{eq:2} and \eqref{eq:17} brings about the same results as the
solution of eq.\eqref{eq:2} does (see eq.\eqref{eq:8}), but with the
renormalized transition probabilities induced by the phonon
bottleneck:

\begin{equation}\label{eq:18}
    W_{m\rightarrow n}^{(r)}=W_{m\rightarrow n}\left[1+\frac{2\pi^2v^3\tau_{ph}(\omega_{mn})Nw_{mn}|\rho_{0m}-\rho_{0n}|}
    {3\omega^2_{mn}\Delta\omega_{mn}[1+i\omega\tau_{ph}(\omega_{mn})]}\right]^{-1}
\end{equation}

To calculate the susceptibility, we need to know, first of all,
elements of the relaxation matrix $\mathbf{W}$, i.e., the transition
probabilities for each pair of states of a paramagnetic ion induced
by the electron-phonon interaction. At the next step, the parameters
of the cross-relaxation rates and the phonon lifetimes introduced in
expressions \eqref{eq:14} and \eqref{eq:17}, respectively, may be
determined from a comparison of the calculated real and imaginary
parts of the susceptibility \eqref{eq:8} with the experimental data.

\section{Spectral properties of isolated $\mathrm{Ho^{3+}}$ ions in $\mathrm{LiYF_4}$}

The crystal lattice of LiYF$_4$ has a space group C$_{4h}^6$ with
the lattice constants $a$ = 0.5164 nm, $c$ = 1.0741 nm.
\cite{garcia1993} The Ho$^{3+}$ ions substitute for Y$^{3+}$ ions in
the sites with the point symmetry S$_4$, coordinates of the four
nearest fluorine ions at the distance $R_1$=0.2244 nm relative to
the Ho$^{3+}$ ion in the site (000) equal (x y z), (-x -y z), (x -y
-z), (-x y -z), where x = (t-1/2)$a$, y = (1/2-p)$a$, z = -q$c$, and
p=0.2817; t=0.1645; q=0.0813. The next nearest fluorine ions are at
the distance $R_2$=0.2297 nm, they form a deformed tetrahedron as
well with x = t$a$, y = (1/2-p)$a$, z = (q-1/4)$c$).

\begin{figure}[htbp]
\includegraphics*[bb=15 15 188 279,scale=1,clip]{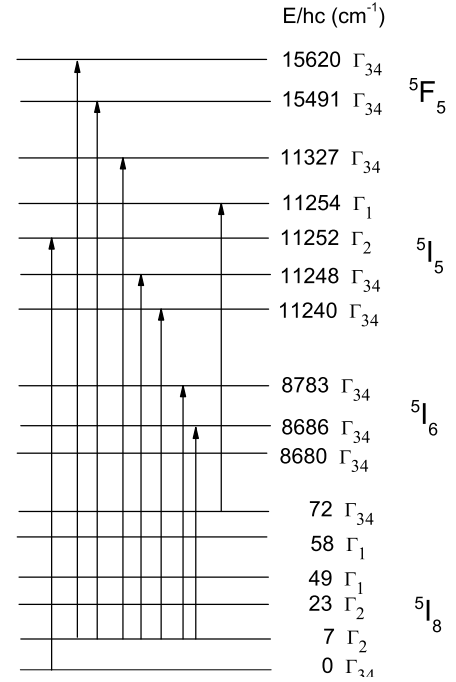}
\hspace{-12pt} \caption{Optical singlet-doublet transitions in the
absorption spectrum of impurity Ho$^{3+}$ ions in LiYF$_4$ explored
in the piezospectroscopic measurements.} \label{fig:1}
\end{figure}

The energy level pattern of the electronic 4f$^{10}$ configuration
of the Ho$^{3+}$ ion in LiYF$_4$ was studied in quite a few
works\cite{15Kara,16Agla,17Walsh}, fragments of this pattern are
shown in Fig.\ref{fig:1}. The spectrum consists of singlets
($\Gamma_1$ and $\Gamma_2$) and non-Kramers doublets (
$\Gamma_{34}$) corresponding to irreducible representations
$\Gamma_k$ of the S$_4$ group. Being interested in low temperature
magnetic properties of the system, we can consider only the lower
part of the energy spectrum described by the parameterized single
ion Hamiltonian operating in the space of 136 products
$|LSJJ_z>\otimes|II_z>$ of the electronic and nuclear spin functions
(there is only one Holmium isotope $^{165}$Ho with the nuclear spin
I=7/2, L, S, J are the electronic orbital, spin and total angular
moments, respectively, the lowest electronic multiplet is
$^5$I$_8$):

\begin{equation}\label{eq:19}
    H=H_0+H_{e-ph}\quad;\quad H_0=H_{CF}+H_{hf}+H_{Z}+H_{S}
\end{equation}

Here

\begin{eqnarray}\label{eq:20}
    H_{CF}&=&a_2B_2^0O_2^0+a_4(B_4^0O_4^0+B_4^4O_4^4+B_{4}^{-4}O_{4}^{-4})\nonumber\\
    &&+a_6(B_6^0O_6^0+B_6^4O_6^4+B_{6}^{-4}O_{6}^{-4})
\end{eqnarray}

is the crystal field Hamiltonian ( $B_p^k$ are the crystal field
parameters, $O_p^{|k|}$  and $O_p^{-|k|}$ are the real and imaginary
Stevens operators, respectively; $a_2 = \alpha = -1/450$, $a_4 =
\beta = -1/30030$, $a_6 = \gamma =  -5/(189.143^2)$ are the reduced
matrix elements of the Stevens operators in the manifold of the pure
$^5$I$_8$ states), $H_{hf}$ is the Hamiltonian of the magnetic
hyperfine interaction (the electrostatic quadrupole interaction can
be neglected), $H_Z=g_J\mu_B\mathbf{JB}$ is the electronic Zeeman
energy in the magnetic field $\mathbf{B}$; the last two terms in
\eqref{eq:19}, the Hamiltonian of the electron-phonon interaction
$H_{e-ph}$ and the energy of interaction with random lattice strains
$H_S$, are specified below.
\begin{figure}[htbp]
\includegraphics*[bb=15 15 360 250,scale=0.9,clip]{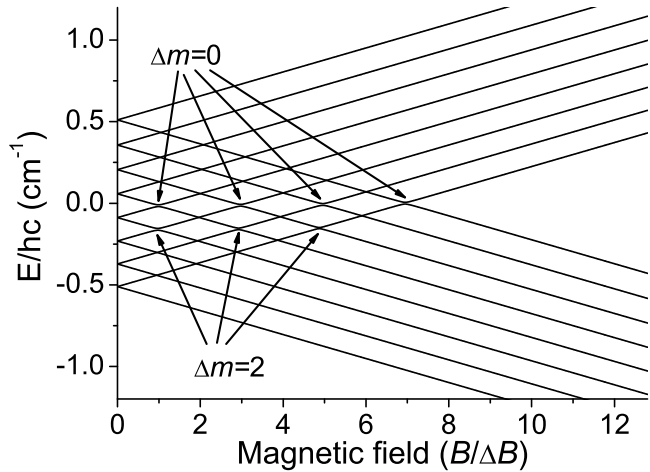}
\hspace{-12pt} \caption{Electron-nuclear sublevels of the Ho$^{3+}$
ground doublet $^5$I$_8$($\Gamma_{ 34}$) in the magnetic field
$\mathbf{B} || c$. The magnetic field strength is given in units of
$\Delta B =$ 23.74 mT.} \label{fig:2}
\end{figure}
As it was shown in [\onlinecite{shakurov2005}], the submillimeter
EPR spectra of Ho$^{3+}$ ions in LiYF$_4$ are well described by the
effective \emph{g}-factor $g_J$ =1.21 (slightly different from the
value 5/4 for the pure $^5$I$_8$ multiplet), the hyperfine constant
$A$ = 0.795 GHz, and the crystal field parameters given in Table 1.
\begin{table} \label{tab:1}
\begin{tabular}{|c c|c|c|c|c|}
  \hline

  p&k&\multicolumn{2}{|c|} {$B_p^k$}&$B_p^k(A_g^1)$&$B_p^k(A_g^2)$\\ \cline{3-4}
  \multicolumn{2}{|c|}{}& Exp. & Theo.& &  \\ \hline
  \multicolumn{2}{|c|}{1}&2&3&4&5\\ \hline
  2&0&190.4&154&603&-891\\ \hline
  4&0&-78.2&-89&125&718\\ \hline
  4&4&-657.2&-700&2397&1434\\ \hline
  4&-4&-568.6&-613&3700&717\\ \hline
  6&0&-3.2&-2.1&113&-416\\ \hline
  6&4&-364.0&-322&937&-843\\ \hline
  6&-4&-222.3&-265&1738&-889\\ \hline
  \hline
p&k&\multicolumn{2}{|c|} {$B_p^k(B_g^1)$}&\multicolumn{2}{|c|} {$B_p^k(B_g^2)$}\\
\cline{3-6}
  \multicolumn{2}{|c|}{}& Exp. & Theo.& Exp. & Theo.  \\ \hline
  \multicolumn{2}{|c|}{6}&7&8&9&10\\ \hline
  2&2&1644&1800&3814&3590\\ \hline
  2&-2&1846&2070&-836&-1620\\ \hline
  4&2&-454&-780&-1532&-1980\\ \hline
  4&-2&1885&3660&1424&1900\\ \hline
  6&2&188&230&-243&-730\\ \hline
  6&-2&-543&-520&-658&-710\\ \hline
  6&6&-858&-740&-1444&-1750\\ \hline
  6&-6&-738&-990&-1245&-2010\\ \hline

\end{tabular}

\begin{tabular}{|c c|c|c||c c|c|c|}
\hline
  p&k& $B_{pk,1}(E_g)$ & $B_{pk,2}(E_g)$ &  p & k & $B_{pk,1}(E_g)$
  &$B_{pk,1}(E_g)$\\ \hline
  \multicolumn{2}{|c|}{11}&12&13&\multicolumn{2}{|c|}{14}&15&16 \\
  \hline
  2&1& 1951 & 3595 & 2 & -1 & -3595& 1951\\ \hline
  4&1& -2056 & -1996 & 4 & -1 & 1996& -2056\\ \hline
  6&1& -1198 & -256 & 6 & -1 & 256& -1198\\ \hline
  4&3& 23372 & 16638 & 4 & -3 & 16638& -23372\\ \hline
  6&3& 2093 & -1674 & 6 & -3 & -1674& -2093\\ \hline
  6&5& -7558 & -600 & 6 & -5 & 600 & -7558\\ \hline
\end{tabular}
\caption{Crystal field parameters and electron-deformation coupling
constants (cm$^{-1}$)}
\end{table}

It should be noted that this set of the crystal field parameters is
related to the crystallographic system of coordinates (the
quantization axis z is parallel to the crystal c-axis). This
statement is based on the earlier studies of the piezo-spectroscopic
and nonlinear Zeeman effects in LiYF$_4$:Tm$^{3+}$
[\onlinecite{18Vino,19Yu}], and the anisotropic parastriction in the
concentrated paramagnets LiTmF$_4$ \cite{20Al} and LiErF$_4$
\cite{21Buma}. The calculated hyperfine structure of the ground
electronic doublet in an external magnetic field directed along the
crystal symmetry axis is shown in Fig.\ref{fig:2}. The spectrum
consists of the two nearly equidistant groups of electron-nuclear
sublevels with positive and negative slopes, respectively. The
electron-nuclear sublevels intersect at the magnetic field values $B
\cong|m'+m| \Delta B$, where $m$,$m'$= $I_z$ and, to first order in
$A$, $\Delta B=A/2g_J\mu_B$. We can distinguish odd (C$_{odd}$,
$|m'-m|=\Delta m=2k$) and even (C$_{even}$,$ |m'-m|=\Delta m=$2k+1)
crossing points at magnetic field $B=(2n+1)\Delta B$ and $B=2n\Delta
B$, respectively, (k,n =0,1,2,3). The mixing of the ground
$\Gamma_{34}$ doublet with the excited singlets $\Gamma_1$ and
$\Gamma_2$ by the magnetic hyperfine interaction opens the gaps at
the odd crossing (anti-crossing) points with the maximum values for
$\Delta m=2$ of an order of 0.35 GHz (see Fig.\ref{fig:2}). However,
in addition to the avoided level crossings with $\Delta m=2$, gaps
of comparable values ($\sim$0.28 GHz) were observed at the odd
crossing points for the electron-nuclear sublevels with $\Delta m=0$
in the EPR spectra of the isotopically enriched sample
$^7$LiYF$_4$:Ho$^{3+}$ (0.1\%) \cite{shakurov2005}. We suppose that
these splittings are induced by random strains due to intrinsic
lattice defects.
As we shall see below, the presence of random strains is confirmed
by some specific features of magnetic field dependences of the
dynamic susceptibility as well, and to account for the corresponding
random crystal field effects, we introduce the Hamiltonian
$H_s=a_2(B_2^2O_2^2+B_2^{-2}O_2^{-2})$ with the initial values of
parameters $B_2^2=\mp$0.4525 cm$^{-1}$; $B_2^{-2}=\pm$0.4752
cm$^{-1}$ which has been already used in [\onlinecite{shakurov2005}]
to describe the $\Delta m=0$ anti-crossings observed in the EPR
spectra of the  sample isotopically enriched in $^7$Li.

\begin{figure}[htbp]
\includegraphics*[bb=15 15 400 230,scale=0.9,clip]{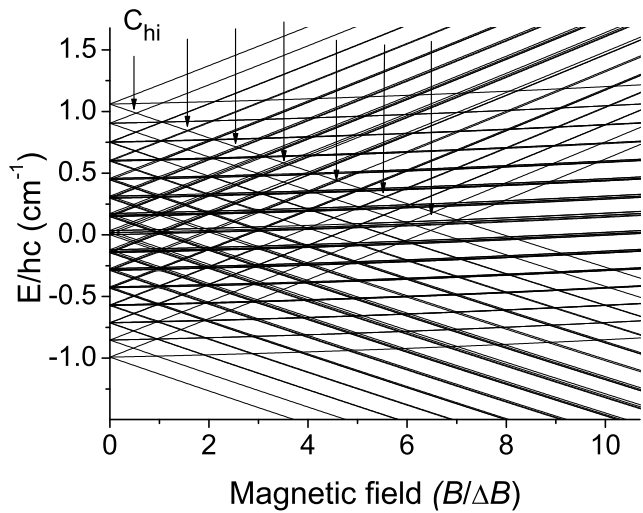}
\hspace{-12pt} \caption{Electron-nuclear sublevels of the ground
state of a pair of distant holmium ions in the magnetic field
$\mathbf{B} || c$. Arrows show half-integer crossings} \label{fig:3}
\end{figure}

At low temperatures interactions between Ho$^{3+}$ ions play an
important role in relaxation processes in the coupled
electron-nuclear subsystem\cite{Giraud2003,Giraud2001,barbara2004}.
For a pair of distant Ho$^{3+}$ ions, the energies of lower 256
electron-nuclear sublevels versus the magnetic field are shown in
Fig.\ref{fig:3}. This figure presents a new plot of the sum of
single ion energies related to a ground  $\Gamma_{34}$ doublet where
the random "rhombic" crystal field is taken into account. It is
described by the Hamiltonian $H_S$ and mainly affects the
electron-nuclear sublevels with the nuclear spin projections $m=\pm
1/2$. The corresponding shifts (about 0.01 cm$^{-1}$) cause
splittings of degenerate energy levels of a dimer which are
concentrated close to the centre of gravity of the spectrum in
Fig.\ref{fig:3} (in particular, in zero magnetic field, 32 states
have zero energy (8 with positive, 8 with negative, and 16 with zero
slope in the magnetic field). It is well seen in Fig.3 that there
are many additional crossings of the electron-nuclear sublevels in
the pair spectrum as compared to the single ion spectrum
(Fig.\ref{fig:2}), and the most important qualitative peculiarity is
the presence of half-integer crossings (C$_{hi}$) at the magnetic
field values $B\cong (n+1/2)\Delta B$ for $n$=0,1...6.

Widths of transitions between the electron-nuclear sublevels of the
ground doublet and the first excited singlet measured in EPR spectra
are strongly dependent on the holmium
concentration\cite{shakurov2005} at liquid helium temperatures.
Assuming a Gaussian distribution, the inhomogeneous broadening can
be defined by the FWHM (full width at half maximum) of 175
MHz\cite{shakurov2005}. According to Ref.[$\cite{23Martin}$],
frequencies of resonance transitions of the two non-equivalent
nearest neighbor fluorine nuclei in the super-hyperfine
$^5$F$_5$-$^5$I$_8$ spectrum equal 29.85 and 36.86 MHz.  The
corresponding widths of Ho$^{3+}$ energy levels due to the
super-hyperfine interactions may be estimated as 130 MHz, and the
total linewidths at low concentrations ($\sim$0.1\%) are close to
300 MHz. The magnetic dipole-dipole interactions dominate at holmium
concentrations larger than 0.5\%, and the observed line widths
exceed 1000 MHz in the sample containing about 1\% of impurity
Ho$^{3+}$ ions.

\section{Modeling of the electron-phonon interaction and spin-lattice relaxation rates}

We shall consider interactions of impurity Ho$^{3+}$ ions with
acoustical phonons with energies less than 40 K at liquid helium
temperatures (energies of optical phonons in LiYF$_4$  are larger
than 100 K\cite{24Sal}). The corresponding phonon wave lengths
exceed essentially the lattice constants, and we can use the lattice
elasticity theory to describe corresponding dynamic lattice
deformations.

In the linear approximation, the Hamiltonian of the electron-phonon
interaction corresponding to the modulation of the crystal field by
elastic waves can be written as \cite{25Mal}
\begin{eqnarray}\label{eq:23}
    H_{e-ph}&=&\sum_{\Gamma j\lambda}\sum_{pk}B_{p,\lambda}
^k(\Gamma^j)e_\lambda(\Gamma^j)a_pO_p^k\\
&&+i\left.\left.\sum_\alpha\vartheta_\alpha\right[H_{CF}+H_{hf}+H_{Z},(J_\alpha+I_\alpha)\right]\nonumber
\end{eqnarray}
here the first and second terms on the right describe the
electron-deformational and electron-rotational interactions,
respectively. Linear combinations $e_\lambda(\Gamma^j)$ of the
deformation tensor components ($e_{\alpha\beta}$) which transform
according to irreducible representations of the factor group
C$_{4h}$ of the lattice ($e(A_g^1)=e_{zz}$;
$e(A_g^2)=(e_{xx}+e_{yy})/2$; $e(B_g^1)=e_{xx}-e_{yy}$;
$e(B_g^2)=e_{xy}$; $e_1(E_g)=e_{xz}$; $e_2(E_g)=e_{yz}$) and the
rotation vector $\mathbf{\vartheta}$  are linear in the phonon
annihilation ($a_{jq}$) and creation ($a_{jq}^+$) operators. In
particular,
\begin{equation}\label{eq:22}
    e_{\alpha\beta}=\sum_{q,j=j_{ac}}\frac{q}{\sqrt{Nm}}[\alpha\beta,1]\left(\frac{\hbar}{2\omega_j(\mathbf{q})}\right)^{1/2}
    (a_{j\mathbf{q}}+a_{j-\mathbf{q}}^+)
\end{equation}
where
\begin{equation}\label{eq:30}
    [\alpha\beta,\sigma]=\frac{1}{2}\left(e_\alpha(j\mathbf{q}_0)q_{0\beta}+\sigma
    e_\beta(j\mathbf{q}_0)q_{0\alpha}\right), \sigma=\pm1
\end{equation}
the sum is taken over acoustical branches $j_{ac}$ of the lattice
vibrational spectrum, $N$ is the number of unit cells having the
mass $m$, $e_\alpha (j_{ac}\mathbf{q}_0)$ are the components of the
unit polarization vector in the elastic wave with the unit
propagation vector $\mathbf{q}_0=\mathbf{q}/q$ and the frequency
$\omega_j(\mathbf{q})$. The components of the rotation vector
$\theta_\gamma$ are given by the same expression \eqref{eq:22} where
$[\alpha\beta , -1]$ is substituted for $[\alpha\beta , 1]$  and
$\alpha\neq\beta\neq\gamma$. Parameters of the
electron-deformational interaction
\begin{eqnarray}\label{eq:24}
\left\{
  \begin{array}{ll}
    B_p^k(A_g^1)=&B_{p,zz}^k  \\
    B_p^k(A_g^2)=&B_{p,xx}^k+B_{p,yy}^k \\
    B_p^k(B_g^1)=&(B_{p,xx}^k-B_{p,yy}^k)/2\\
    B_p^k(B_g^2)=&2B_{p,xy}^k\\
    B_{p,1}^k(E_g)=&2B_{p,xz}^k\\
    B_{p,2}^k(E_g)=&2B_{p,yz}^k\\
    \end{array}
\right.
\end{eqnarray}
can be computed if the crystal field parameters $B_p^h$ for an
impurity rare earth ion are known as explicit functions of
coordinates $X_{\lambda L,\alpha}$, of the host lattice ions (the
vector $\mathbf{R}_{\lambda L}$ defines the equilibrium position of
an ion $\lambda$ in the unit cell $L$). The corresponding
calculations were carried out in the framework of the exchange
charge model of the crystal field\cite{25Mal}. Parameters of the
crystal field Hamiltonian are represented by a sum of two terms
\begin{eqnarray}\label{eq:27}
    B_p^k&=&e^2K_p^k\sum_{\lambda L}\left[-Z_\lambda(1-\sigma_p)\frac{<r^p>}{R_{\lambda
    L}^{p+1}}\right.\nonumber\\
    &&+\left.\frac{2(2p+1)}{7R_{\lambda L}}S_p(R_{\lambda L})\right]
    O_{p}^{k}(\vartheta_{\lambda L},\varphi_{\lambda L})
\end{eqnarray}
related to the electrostatic fields of point lattice ions with the
effective charges $eZ_\lambda$  and "exchange" charges at the
neighbor ions proportional to the quadratic forms of the overlap
integrals of the 4f-electron and ligand wave functions $S_s =
<4f0|n''s0>$, $S_\sigma  = <4f0|n''p0>$, $S_\pi  = <4f1|n''p1>$
(only outer filled $n''s^2$ and $n''p^6$ electronic shells of ligand
ions are considered):
\begin{equation}\label{eq:28}
    S_p(R)=G_sS_s^2(R)+G_\sigma
    S_\sigma^2(R)+(2-\frac{p(p+1)}{12})G_\pi S_\pi^2(R)
\end{equation}
Here $K_p^k$ are the numerical coefficients\cite{25Mal}, $\sigma_p$
are the shielding constants, $<r^p>$ are the moments of the 4f
electron charge density, $O_p^k(\vartheta,\varphi)$ are the
homogeneous Stevens polynomials formed from the direction cosines of
ligand radius-vectors (the spherical coordinates of a ligand are
$R,\vartheta,\varphi$ relative to the rare earth ion at the origin).
Calculations were performed with  $\sigma_2$ = 0.579,  $\sigma_4
=\sigma_6 = 0$\cite{26Erdos}, $<r^2>$ = 0.695, $<r^4>$ = 1.219,
$<r^6>$ = 4.502 (atomic units)\cite{27Free}, the lattice sums were
computed by the Ewald method, the dependences of the overlap
integrals (computed with the radial wave functions from
Refs.[\onlinecite{28Sov,29Cle}]) on the distance $R$ (in atomic
units) between the ions were approximated by functions
$S_0\exp(-bR)$ where $S_0$=2.2298, 0.6535, 1.3239; $b$=1.2554,
0.88259, 1.1596  for $s$,$\sigma$ and $\pi$ bonds, respectively. The
values of the model parameters $G_s = G_\sigma = 5.6$, $G_\pi  =
2.85$ were obtained from fitting the calculated crystal field
energies to the experimental data. The model results in the set of
crystal field parameters (Table 1, column 3) which are well
comparable with those found from the analysis of the optical and EPR
spectra (Table 1, column 2)\cite{shakurov2005}. The coupling
constants $B^k_{p,\lambda}(\Gamma^j)$  in the Hamiltonian of the
electron-deformational interaction presented in Table 1 (columns
4,5,8,10,12,13,15,16) were calculated using the same parameters of
the model.

Now we have in hands all the data which are necessary to calculate
the single phonon transition probabilities introduced in section 2:

\begin{eqnarray}\label{eq:29}
    &&W_{m\rightarrow f}=\frac{\omega^3_{mf}}{\pi\hbar
    \rho}\sum_{j=j_{ac}}\int\frac{\sin\theta d\theta d\varphi}{4\pi
    v_j^5(\theta\varphi)}\times\\
    &&\left|\sum_{\alpha\beta}\right. <f|\sum_{pk}B^k_{p,\alpha\beta}[\alpha\beta,+1]a_pO_p^k\nonumber\\
    &&+    i\hbar\omega_{fm}[\alpha\beta,-1]\left.\sum_\gamma\varepsilon_{\alpha\beta\gamma}(J_\gamma+I_\gamma)|m>\right|^2
    (n_0(\omega_{mf})+1)\nonumber
\end{eqnarray}
Here $\varepsilon_{\alpha\beta\gamma}$ is the unit antisymmetric
tensor, $v_j(\theta\varphi )$ is the sound velocity in the direction
of the phonon wave vector $\mathbf{q}$ determined by the angular
coordinates $\theta$ and $\varphi$ . The integrals
$\sum_{j=j_\alpha}\int[\alpha\beta,\sigma][\gamma\delta,\sigma']d\Omega/4\pi
v^5_{j\alpha}$ were computed in Ref.[\onlinecite{30Anti}].

\begin{figure}[htbp]
\includegraphics*[bb=10 10 300 400,scale=1,clip]{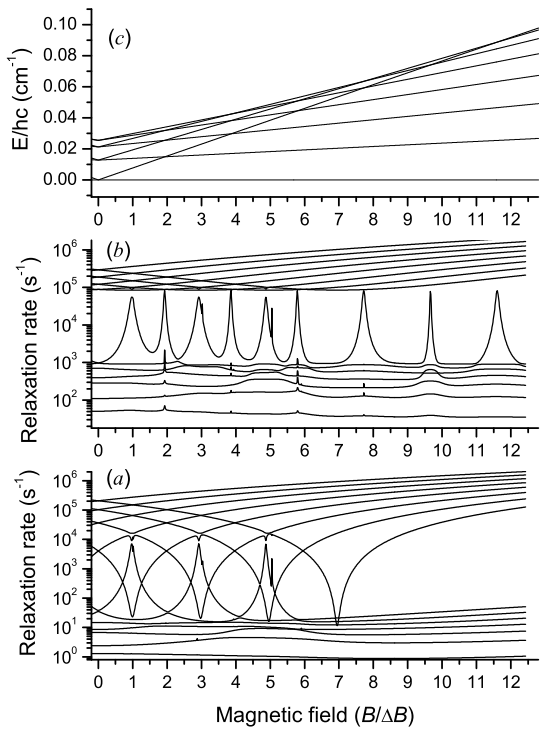}
\hspace{-12pt} \caption{Electron-phonon relaxation rates of a single
Ho$^{3+}$ ion in the phonon bath at the temperature 2 K. (a) -
calculated eigenvalues of the relaxation matrix defined in the space
of 16 electron-nuclear sublevels of the ground doublet, (b) -
calculated lower relaxation rates corresponding to one-phonon
transitions between all sublevels of the ground doublet and the
first singlet, (c) - relative energies of the electron-nuclear
sublevels of the first excited singlet in the magnetic field
$\mathbf{B}|| c$.} \label{fig:4}
\end{figure}

The analysis of the magnetic field and temperature dependences of
the relaxation rates involved the numerical diagonalization of the
matrix of the Hamiltonian $H_0$ (see eq.\eqref{eq:19}) for the fixed
values of the magnetic field (the matrix was constructed in the
space of 136 functions $|^5I_8,JJ_z>\otimes|II_z>$ of the lowest
multiplet), calculations of transition probabilities for each pair
of the electron-nuclear sublevels using the corresponding
eigenfunctions of $H_0$, and the diagonalization of the relaxation
matrix $\mathbf{W}$. Fig.\ref{fig:4} (a) shows the magnetic field
dependences at the temperature 2 K of the eigenvalues of the
relaxation matrix defined in the subspace of sublevels of the ground
doublet. In this case transitions are mainly induced by the dynamic
deformations of $B_g$ symmetry between the sublevels with the same
projections of the nuclear spin. However, mixing of wave functions
of the ground doublet with the wave functions of the excited
singlets allows transitions with the nuclear spin reverting as well.
At high magnetic fields ($B>$0.18 T), there are two sets of
solutions each consisting of eight branches corresponding to allowed
(fast relaxation rates exceeding $10^3 s^{-1}$) and forbidden (slow
relaxation rates less than $10^2 s^{-1}$) transitions (of course,
there is always a zero solution corresponding to the equilibrium
state of the system). At lower magnetic fields we see specific
variations of four upper branches at the  $\Delta m=0$
anti-crossings and of three lower branches at the  $\Delta m=2$
anti-crossings pointed in Fig.\ref{fig:2}. The additional narrow
peaks superimposed on the broad peaks are caused by more narrow
$\Delta m=-2$ anti-crossings which are shifted from the $\Delta m=2$
anti-crossings along the magnetic field axis because the spectrum is
only approximately equidistant.

Fig.\ref{fig:4}(b) demonstrates a remarkably different behavior of
the considered single ion relaxation rates at the same temperature,
2 K, when the transitions between the first excited singlet and the
ground doublet are taken into account (these transitions are induced
by lattice deformations of $E_g$ symmetry). There are again two sets
of fast and slow relaxation rates, but all rates become larger by
one-two orders of magnitude, the dips at the $\Delta m=0$
anti-crossings disappear, and six additional peaks in the lower
branches are observed due to the anti-crossings between the
sublevels of the first excited singlet  $\Gamma_2$ which are shown
in Fig.\ref{fig:4}(c). It follows from the results of calculations
of the relaxation rates at liquid helium temperatures that the
frequency dependence of the out-of-phase susceptibility of Ho$^{3+}$
ions should exhibit a maximum at frequencies close to $10^5-10^6$ Hz
contrary to the earlier observations\cite{Giraud2003} of this
maximum at frequencies below $10^3$ Hz. This contradiction
stimulated us to test the theory by direct measurements of the
electron-phonon coupling constants in the piezospectroscopic
experiments described in the following section.

\section{Expérimental}
\subsection{Piezospectroscopic measurements}

To determine parameters of the Hamiltonian \eqref{eq:23}
corresponding to the interaction of the Ho$^{3+}$ ion with lattice
strains of $B_g$-symmetry, we measured splittings of nine
$\Gamma_{34}$ doublets (see Fig.\ref{fig:1}) induced by the uniaxial
stress applied in the basis plane of the lattice. The sample
LiYF$_4$ containing 1 at.\% of holmium was grown by the
Bridgman-Stockebarger method and oriented with the X-ray
diffractometer. Experiments were performed at temperatures 4.2 K and
77 K, the pressure $p$ up to 220 MPa was supplied along three
different directions in the (001) plane. Splittings of the optical
lines corresponding to singlet-doublet transitions in the $\sigma$
polarized absorption spectra (the wave vector of the incident light
was parallel to the c-axis) were measured directly or from the
linear dichroism signals\cite{18Vino}.

The uniaxial pressure in the plane (001) at the angle $\varphi$
relative to the $a$ axis induces the following nonzero components of
the deformation tensor:
\begin{eqnarray*}
\left\{
  \begin{array}{ll}
    e(A_g^1)=&-S_{13}p  \\
    e(A_g^2)=&- (S_{11}+S_{12})p/2\\
    e(B_g^1)=&-[(S_{11}-S_{12})\cos2\varphi+S_{16}\sin2\varphi]p\\
    e(B_g^2)=&-(2S_{16}\cos2\varphi+S_{66}\sin2\varphi) p/4
  \end{array}
\right.
\end{eqnarray*}
Here $S_{ij}$ are the components of the elastic compliance tensor of
LiYF$_4$ measured in Ref.[\onlinecite{31Blan}], in particular,
$S_{11}-S_{12}= 18.6$, $S_{16}= 8.01$, $S_{66}= 57.7$ ($10^{-6}
/MPa$). In the linear approximation, the corresponding splitting of
the doublet with the wave functions $|\Gamma_{34}\pm>$ equals
\begin{eqnarray}\label{eq:31}
\Delta(\Gamma_{34},\varphi)&=&2\left|\left<\Gamma_{34}+\right|\sum_{pk,i=1,2}B_p^k(B_g^i)e(B_g^i)a_pO_p^k\left|\Gamma_{34}-\right>\right|\nonumber\\
&=& p(C+D\cos(4\varphi-\varphi_0))^{1/2}
\end{eqnarray}

Using the data obtained at three different values of the angle
$\varphi$  (see Table 2), we determined for each doublet the
corresponding deformation potentials $C$ and $D$ and the directions
$\varphi_0$ of the petals of the four-fold rosettes \eqref{eq:31}.
Then the set of 16 electron-deformation parameters $B_p^k(B_g^i)$
($p$ $k\rightarrow$ 2 2, 2 -2, 4 2, 4 -2, 6 2, 6 -2, 6 6, 6 -6) was
varied starting from the values obtained in the framework of the
exchange charge model (Table 1, columns 8,10) to obtain the best
estimations of the measured splittings. Results of the fitting
procedure are given in Table 1 (columns 7,9), and the calculated and
experimental splittings are compared in Table 2. We had only 18
independent nonlinear equations for 16 variables. It should be noted
that the positive direction of the b-axis relative to the a-axis was
unknown, and signs of angles $\varphi$ were checked in the fitting
procedure. Possible errors in the measured splittings (in
particular, due to random internal strains and errors in the
orientation of the sample) are rather large ( $\pm$0.15
cm$^{-1}$/100 MPa), the linear approximation for some closely spaced
doublets (having numbers 3,5,6 in Table 2) is not valid, and the
parameters $B_p^k(B_g^i)$ have not been determined unambiguously.
However, as it is seen in Table 1, the final set of parameters
$B_p^k(B_g^i)$ does not differ qualitatively from the starting
values. Thus, the model used in calculations of the parameters of
the electron-phonon interaction has been additionally approved. A
possible error in the estimations of the matrix elements of
electronic operators in the Hamiltonian of electron-phonon
interaction \eqref{eq:23} which we need to calculate the transition
probabilities between the sublevels of the ground doublet does not
exceed 24 \%.

\begin{table}\label{Tab:2}
\begin{tabular}{|c|c c|c c|c c|c c|}
  \hline
  N & \multicolumn{2}{|c|}{Energy of $\Gamma_{34}$} & \multicolumn{6}{|c|}{Splitting $\Delta(\varphi)$ (cm$^{-1}$/100MPa)} \\
   \cline{4-9}& \multicolumn{2}{|c|}{(cm$^{-1})$} & \multicolumn{2}{|c|}{$\varphi=5^\circ$} & \multicolumn{2}{|c|}{$\varphi=30.7^\circ$}  & \multicolumn{2}{|c|}{$\varphi=45^\circ$}
   \\ \hline
   1&$^5$I$_8$& 0 & 0.33 & (0.43) & 0.67 & (0.63) & 0.53 & (0.53)\\
   2&$^5$I$_8$& 72 & 1.12 & (1.50) & 1.50 & (1.98) &1.14 & (1.57)\\
   3&$^5$I$_6$& 8686 & 0.62 & (0.37) & 0.71 & (0.35) & 0.38 & (0.21)\\
   4&$^5$I$_6$& 8783 & 1.90 & (1.43) & 2.20 & (1.93) & 2.00 & (1.56)\\
   5&$^5$I$_5$& 11240 & 0.34 & (0.49) & 0.44 & (0.45) & 0.34 & (0.30)\\
   6&$^5$I$_5$& 11248 & 0.43 & (0.53) & 0.50 & (0.78) & 0.53 & (0.68)\\
   7&$^5$I$_5$& 11327 & 1.45 & (1.33) & 1.95 & (1.77) & 1.60 & (1.42)\\
   8&$^5$F$_5$& 15491 & 0.52 & (0.72) & 0.67 & (0.90 & 0.62 & (0.85)\\
   9&$^5$F$_5$& 15620 & 1.01 & (1.02) & 1.18 & (1.33) & 0.82 & (1.05)\\
  \hline
\end{tabular}\caption{Splittings of non-Kramers doublets in LiYF$_4$:Ho$^{3+}$ (1\%)
crystals axially compressed in the basis plane. Results of
calculations are in brackets. }
\end{table}

\subsection{Measurements of the dynamic susceptibility}

\begin{figure}[htbp]
\includegraphics*[bb=20 15 186 211,width=0.45\textwidth,clip]{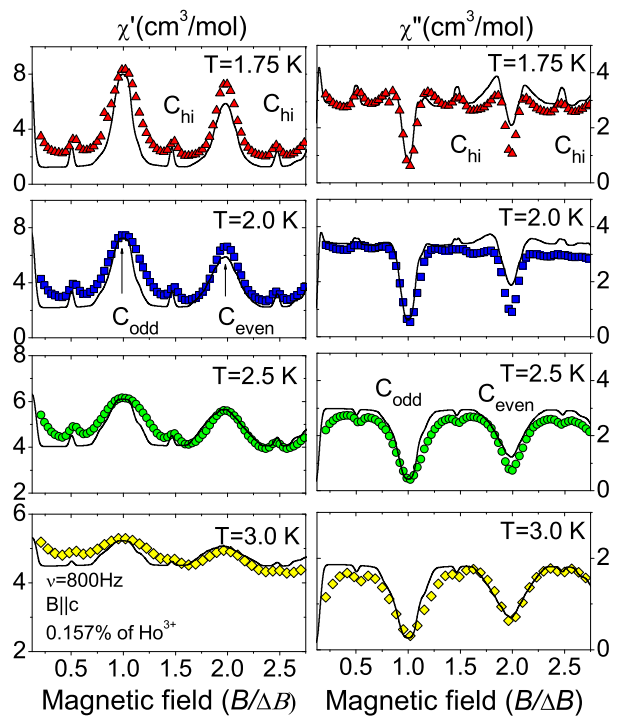}
\hspace{-12pt} \caption{(Color online) Measured and simulated
magnetic field dependences of the ac-susceptibility in
LiYF$_4$:Ho$^{3+}$ (0.157 \%) at different temperatures ( $\nu$= 800
Hz, $\mathbf{B} || c$).} \label{fig:5}
\end{figure}

Earlier the dynamic susceptibility of diluted LiY$_{1-c}$Ho$_c$F$_4$
crystals with holmium concentrations $c$=46 \%, 16.7 \% and 4.5 \%
was studied in Ref.[\onlinecite{33Reich}-\onlinecite{35Reich}] at
low temperatures with main attention for the spin-glass behavior in
a random dipolar-coupled Ising magnet.

In the present work, the dynamic susceptibility of highly diluted
LiY$_{1-c}$Ho$_c$F$_4$ single crystals was measured with a
conventional SQUID magnetometer at frequencies from 20 to 1200 Hz in
the temperature range 1.75 - 4 K. Three samples with dimensions of
4x4x10 mm$^3$ were studied in collinear ac- and dc-magnetic fields
parallel to the c-axis of amplitude of 4 10$^{-4}$ T and up to 0.25
T respectively. The holmium concentrations in these samples
($c$=0.104 \%; 0.157 \%; 0.27 \%) were determined by a comparison of
the measured low frequency susceptibility with the calculated single
ion static susceptibility. The in- and out-of-phase susceptibility
was measured in a more extended dc field, frequencies, temperatures
and concentrations, than in Ref [\onlinecite{Giraud2003}], where the
results of measurements of the in- and out-of-phase susceptibilities
at frequencies 163 and 800 Hz in the sample with holmium
concentration c=0.04\% were published.

\begin{figure}[htbp]
\includegraphics*[bb=15 15 272 230,width=0.45\textwidth,clip]{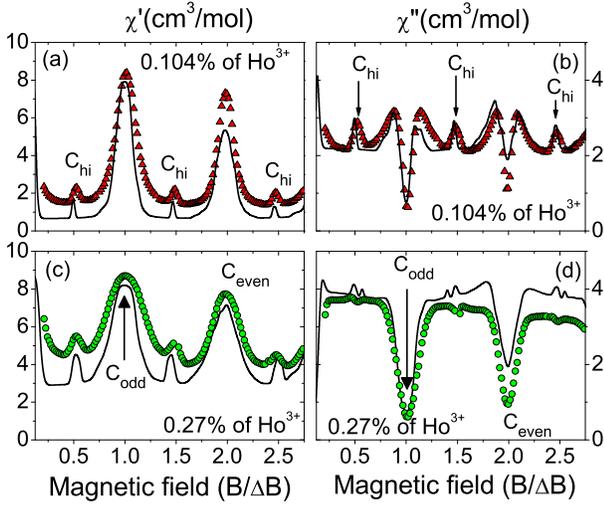}\\
\hspace{-12pt} \caption{(Color online) . Measured and simulated
magnetic field dependences of the ac-susceptibility at the
temperature 1.75 K in LiYF$_4$:Ho$^{3+}$ samples with different
holmium concentrations x (a,b - x = 0.104 \%, c,d - x = 0.27 \%,
$\nu=$ 800 Hz, $\mathbf{B} || c$).} \label{fig:6}
\end{figure}

\begin{figure}[htbp]
\includegraphics*[bb=15 15 201 205,width=0.45\textwidth,clip]{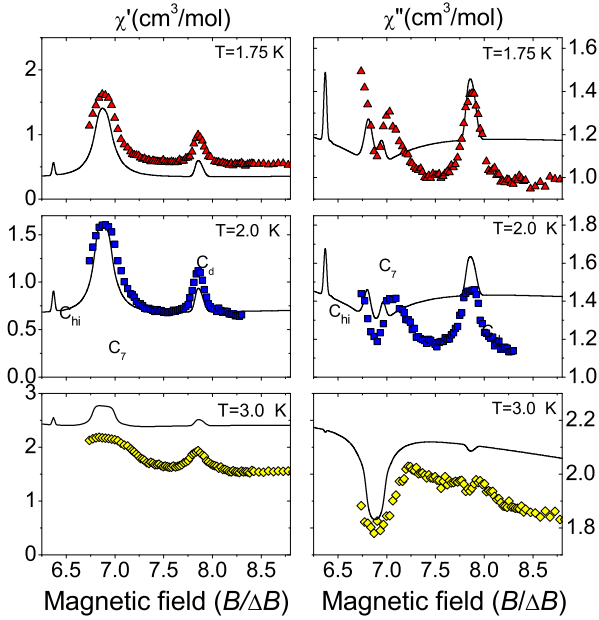}\\
\hspace{-12pt} \caption{(Color online). Measured and simulated
magnetic field dependences of the ac-susceptibility in
LiYF$_4$:Ho$^{3+}$ (0.104 \%) in the region of high-field crossings
at different temperatures ($\nu$ = 1200 Hz, $\mathbf{B} || c$).}
\label{fig:7}
\end{figure}

\begin{figure}[htbp]
\includegraphics*[bb=10 10 209 275,width=0.4\textwidth,clip]{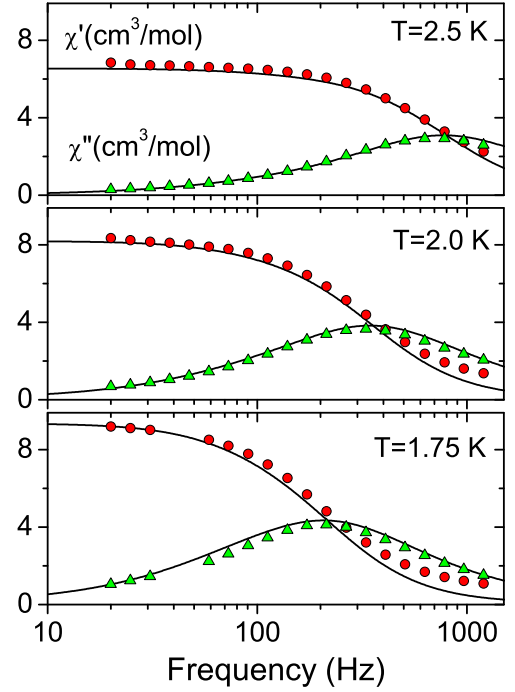}\\
\hspace{-12pt} \caption{(Color online). Measured and simulated
frequency dependences of the dynamic susceptibility in
LiYF$_4$:Ho$^{3+}$ (0.104 \%) at different temperatures ($\mathbf{B}
|| c$, B = 38.5 mT).} \label{fig:8}
\end{figure}

\begin{figure}[htbp]
\includegraphics*[bb=10 10 225 269,width=0.4\textwidth,clip]{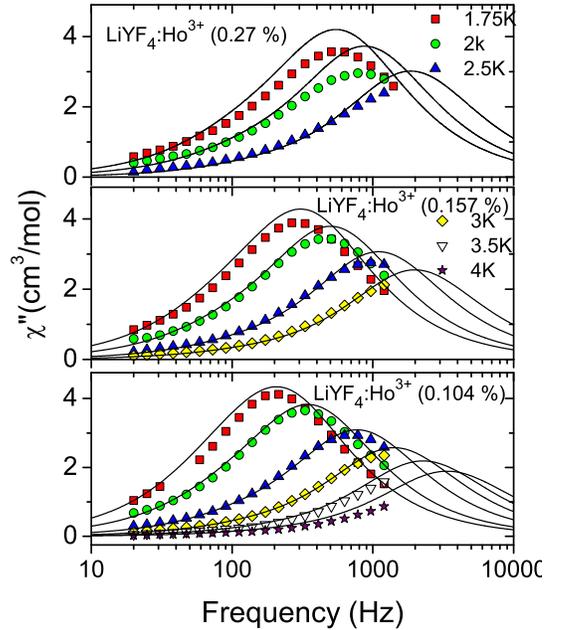}\\
\hspace{-12pt} \caption{(Color online). Measured and simulated
frequency dependences of the out-of-phase susceptibility in
LiYF$_4$:Ho$^{3+}$ samples at different temperatures ($\mathbf{B} ||
c$, B = 38.5 mT).} \label{fig:9}
\end{figure}

\begin{figure}[htbp]
\includegraphics*[bb=10 10 214 297,width=0.4\textwidth,clip]{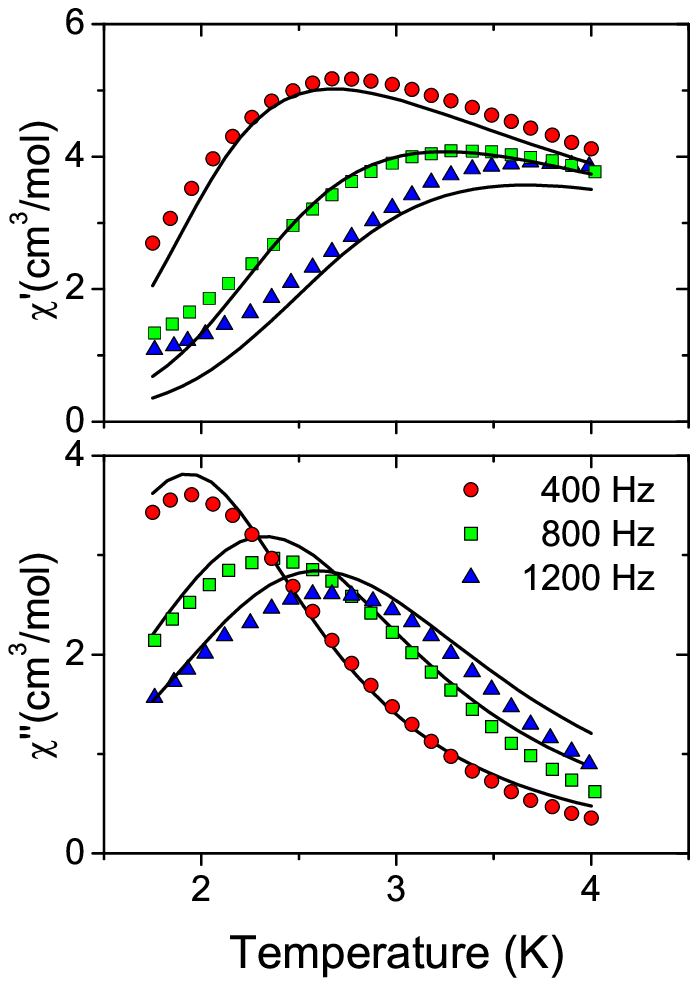}\\
\hspace{-12pt} \caption{(Color online). Measured (symbols) and
simulated (solid curves) temperature dependences of the dynamic
susceptibility in LiYF$_4$:Ho$^{3+}$ (0.104 \%) at different
frequencies of the ac magnetic field ($\mathbf{B} || c$, B = 38.5
mT).} \label{fig:10}
\end{figure}

The results presented in Figs.\ref{fig:5} to \ref{fig:7} show the
non monotonous behavior with well pronounced peaks and dips at
different crossing points of the electron-nuclear energy levels. In
particular, the out-of-phase susceptibility measured in the vicinity
of the crossing points for two holmium concentrations show an
inversion of the sign of  C$_{hi}$ peaks (dips) with respect to the
background (Fig.\ref{fig:6}). Fig.\ref{fig:8} presents the frequency
dependences of the in- and out-of-phase susceptibilities at three
different temperatures while the temperature and concentration
shifts of the out-of-phase susceptibility with frequency are given
in Fig.\ref{fig:9}. Note that in both cases the dc magnetic field
was taken in between two single-ion electron-nuclear crossing points
(38.5 mT, corresponding to 1.62$\Delta B$). The temperature
dependences of the in- and out-of-phase susceptibilities at
different ac frequencies are shown in Fig.\ref{fig:10}.

\section{Simulations of the dynamic susceptibility in $\mathrm{LiYF_4:Ho^{3+}}$ and discussion}

The experimental data presented in Figs.\ref{fig:5}-\ref{fig:10} are
accompanied by results of the corresponding simulations.
Calculations of the susceptibilities involved the numerical
diagonalization of the single ion Hamiltonian $H_0$ (see
eq.\eqref{eq:19}) in the space of 136 electron-nuclear states of the
lowest $^5I_8$ multiplet for the fixed values of the external
magnetic field, the computing of the matrix elements of electronic
operators which are necessary to construct the relaxation matrices
\eqref{eq:14} and \eqref{eq:29}, and at the last step
$\chi'(\omega)=\mathrm{Re}\chi_{zz}(\omega )$ and
$\chi''(\omega)=\mathrm{Im}\chi_{zz}(\omega )$ were obtained at
different frequencies and temperatures using the expression
\eqref{eq:8}.

First of all, let us consider the frequency dependences of the
out-of-phase susceptibilities in Fig. \ref{fig:9} measured at the dc
magnetic field value of 1.62$\Delta B$, which is far enough from the
neighboring single-ion crossing points. The maximum of the
out-of-phase susceptibility shifts to higher frequencies with the
increasing temperature. These shifts give evidence for the
increasing relaxation rates induced by the electron-phonon
interaction. However, using the one-phonon transition probabilities
calculated with the parameters of the electron-deformational
interaction from Table 1, we found the corresponding maxima at the
frequencies $10^4-10^5$ Hz (see Fig.\ref{fig:4} with the plotted
relaxation times), which are about two orders of magnitude higher
than the experimental values. Even more, Fig.\ref{fig:9} also shows
a concentration dependence of the maximum of the out-of-phase
susceptibility, while it is clear that single-ion electron-phonon
interaction cannot produce any dependence of the relaxation rate on
the concentration of relaxing ions. Thus, to fit the simulated
frequency dependences to the experimental curves, we had to take
into account the phonon bottleneck effect. The concentration
dependence of the renormalized transition probability \eqref{eq:18}
is determined by a factor
$K_{mk}=\tau_{ph}(\omega_{mk})N/\Delta\omega_{mk}$. Using the
average sound velocity $v =3.10^3$ m/s \cite{31Blan}, the number of
paramagnetic ions per unit volume $N=2c/V$ (here $V$ is the unit
cell volume which contains two Y$^{3+}$ sites accessible for holmium
ions, $c$ is the concentration determined from measurements of
$\chi'(\omega\rightarrow 0)$), and the widths of spectral
distributions of the resonant phonons $\Delta\omega_{mk} =
2\gamma_{mk}$ = 300 MHz according to the EPR data on the widths of
the singlet-doublet transitions\cite{shakurov2005}, we found it
necessary to introduce two different phonon lifetimes independent on
temperature to describe the experimental data for the sample with
the concentration $c$=0.104\% (see Figs.\ref{fig:8} and
\ref{fig:9}). Namely, we used $\tau_{ph}(\omega)=\tau_g\sim 1 \mu s$
for resonant phonons with frequencies corresponding to transitions
between the electron-nuclear sublevels of the ground electronic
doublet, and $\tau_{ph}(\omega)=\tau_s\sim 0.1 \mu s$ for phonons
with frequencies corresponding to transitions between the excited
singlet and the ground doublet. Actually, the transitions between
the electron-nuclear sublevels of the ground doublet are narrower
than the transitions between the sublevels of the singlet and
doublet crystal field states, and the lifetimes  $\tau_g$ and
$\tau_s$ may differ less than by order of magnitude. However, to fit
the measured frequency dependences of the susceptibilities in the
samples with higher concentrations of holmium ions, we had to
diminish the factors $K_{mk}$ by 1.48 ($c$=0.157\%) and 2.7
($c$=0.27\%) times as compared with these factors for $c$=0.104\%.
Thus, despite the increase of the concentration of the paramagnetic
ions, the phonon bottleneck effect weakens due to broadening of the
spectral bands of the resonant phonons and possible decrease of the
phonon lifetimes.

\begin{figure}[htbp]
\includegraphics*[bb=10 10 229 288,width=0.45\textwidth,clip]{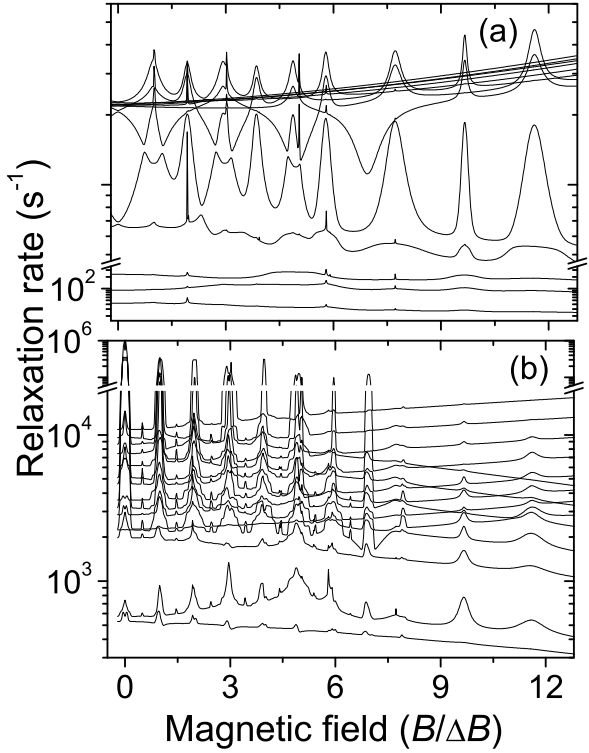}\\
\hspace{-12pt} \caption{The phonon bottleneck and cross relaxation
effects on the relaxation rates. (a) - the relaxation rates at the
temperature 2 K in the sample LiYF$_4$:Ho$^{3+}$ (0.104\%)
calculated with the renormalized electron-phonon transition
probabilities due to the bottleneck effect, (b) - results of
calculations with the cross relaxation terms taken into account.}
\label{fig:11}
\end{figure}

To illustrate the phonon bottleneck effect on the relaxation rates,
we present in Fig. \ref{fig:11}(a) the calculated rates in the
sample with the concentration $c$=0.104\% versus the magnetic field.
These rates are condensed within the much narrower range as compared
with the relaxation rates represented in Figs.\ref{fig:4}(a,b), the
damping effect of the gaps at the $\Delta m=0$ crossings is partly
restored due to the strong suppression of the singlet-doublet
transition probabilities, but there are still no remarkable
variations of the lower branches at the crossing points which may be
expected in accordance with the measured field dependences of the
in- and out-of-phase susceptibilities represented in
Figs.\ref{fig:5}-\ref{fig:7}.

\begin{figure}[htbp]
\includegraphics*[bb=10 10 325 250,width=0.5\textwidth,clip]{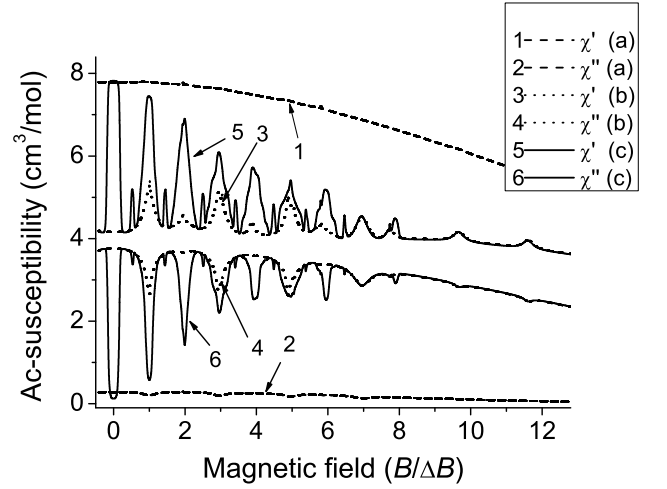}\\
\hspace{-12pt} \caption{ The simulated ac-susceptibility of the
LiYF$_4$:Ho$^{3+}$ (0.27\%) sample at the frequency 800 Hz and the
temperature 2 K, the magnetic field is declined from the c-axis by
1deg. a - the relaxation matrix contains only one-phonon transition
probabilities calculated with the electron-phonon coupling constants
presented in Table 1, b - the transition probabilities are
renormalized due to the phonon bottleneck effect, c - the cross-
relaxation terms are included.} \label{fig:12}
\end{figure}

As it is seen in Fig.\ref{fig:12} the magnetic field dependences of
$\chi'$ and  $\chi''$ obtained without phonon bottleneck and
cross-relaxation effects (with electron-phonon transition
probabilities only, curves 1,2) have nothing in common with the
experimental data (Figs.\ref{fig:5}-\ref{fig:7}). When making use of
electron-phonon transition probabilities renormalized by phonon
bottleneck, we obtain the curves 3,4 showing some overlap with the
measured ones (the specific humps and dips appear at the odd
crossing points) but their "intensities" are lower than the
experimental ones, and also there is no sign of corresponding
features at even and half-integer crossings. However the results
dramatically change when the cross-relaxation is taken into account:
in this case all crossing points manifest themselves properly (the
curves 5,6) and are very close to the experimental results.

The cross-relaxation rates were calculated assuming the Gaussian
line shape
$g^{CR}=(2\pi\Delta)^{-1}\exp[-(\omega_{pn}-\omega_{lm})^2/2\Delta^2]$
of the spectral density of the energy reservoir corresponding to
interactions between the holmium ions. It is enough to consider in
\eqref{eq:14} only three nonzero fitting parameters, $k_{12}$ and
$k_{66}$, which determine the transition probabilities within the
manifold of the electron-nuclear sublevels of the ground doublet,
and $k_{44}$, which determines rates of the singlet-doublet
transitions. For the scaling factor $\delta =2.10^8 s^{-1}$, the
fixed values of $k_{12} = k_{66} =0.1$ and $k_{44}=0.004$
independent on the concentration of the holmium ions were used in
all calculations. However, the dispersion $\Delta$ of frequencies of
the cross-relaxation transitions depends on the concentration, it
was estimated from the EPR linewidths and corrected from a
comparison of the simulated field dependences with the experimental
data ($\Delta$ = 100, 120, 140 MHz for the $g_{44}^{CR}$ line shape,
and $\Delta$ = 185, 200, 240 MHz for the $g_{12}^{CR}$ and
$g_{66}^{CR}$  line shapes for samples with concentrations 0.104,
0.157 and 0.27\%, respectively). As it is seen in
Fig.\ref{fig:11}(b), the cross-relaxation rates play the dominant
role at all crossing points. The most important result of the
cross-relaxation processes is the appearance of the low frequency
($10^2-10^3 s^{-1}$) branch in the spectrum of relaxation rates with
the well resolved maxima at the crossing points. It should be noted
that the calculated field dependences of the relaxation rates may
change remarkably when using another cross-relaxation line shapes
(the Lorentz distributions, in particular)\cite{32Poh}, but to
analyze this problem it is necessary to perform measurements in the
samples with a more wide range of concentrations of paramagnetic
ions.

Small peaks at the half-integer crossings in the field dependences
of $\chi'$ (and the peaks and dips at these points in the field
dependences of $\chi''$) are induced by the cross-relaxation
singlet-doublet transitions (they disappear in the simulated curves
for $k_{44}$=0). The value of the parameter $k_{44}$ used in
simulations (the results are represented in
Figs.\ref{fig:5}-\ref{fig:7}) coincides by an order of magnitude
with the estimation based on the assumption of the prevailing role
of the magnetic dipole-dipole interactions:
$$k_{44}=2\pi c a^6\sum\left(\frac{3xz}{r^5}\right)^2$$
here the sum is taken over all $Y^{3+}$ sites, $a$ is the lattice
constant, and $c$ is the concentration of the impurity holmium ions.
However, values of the parameters $k_{12}$ and $k_{66}$ mentioned
above are much larger than the similar estimations, and it is
possible that the virtual phonon exchange contributes essentially
into the interaction between the holmium ions in the ground state,
in particular, due to strong coupling with the dynamic lattice
deformations of $B_g$ symmetry.

The peaks $C_d$ in the high-field dependences of  $\chi'$ and
$\chi''$ close to the magnetic field value $B=8\Delta B$ in
Fig.\ref{fig:7} correspond to the anti-crossing of the
electron-nuclear sublevels of the excited singlet (see
Fig.\ref{fig:4}). These peaks appear due to a combined action of the
random crystal field (described by the Hamiltonian $H_S$, see
eq.\eqref{eq:19}) and the cross-relaxation processes
(singlet-doublet transitions). So, we had a possibility to determine
independently the $k_{44}$ parameter from fitting intensities of
$C_{hi}$ peaks, and parameters of the Hamiltonian $H_S$ from fitting
intensities of the $C_d$ peaks. The best fit was achieved when the
parameters $B_2^2$, $B_2^{-2}$ from Ref.\onlinecite{shakurov2005}
(see section III above) were increased by a common factor of 1.5,
1.8 and 3.3 for samples studied in this work with holmium
concentrations 0.104\%, 0.157\% and 0.27\%, respectively, and
natural abundances of Li isotopes. It should be noted that we have
not succeeded in exact description of measured real and imaginary
parts of the dynamic susceptibilities at any frequency, temperature
and external magnetic field values. In particular, we obtained
correct positions of the maxima in the frequency dependences of the
out-of phase susceptibilities, but the calculated maximum values are
higher than the measured data (see Fig.9). The discrepancies between
theory and experiment in Fig.7 are the consequences of this lack of
the theory. However, the specific behavior of the susceptibility in
the vicinity of high field anti-crossings is reproduced by
calculations as well. To derive a more elaborated model, we need
additional experimental data at higher frequencies.

 It follows from the
simulations that an additional narrow peak should split out from the
peak $C_d$ in the plot of  $\chi'$ versus the magnetic field (see
the curve 5 in Fig.\ref{fig:12}) in the transverse magnetic field,
in particular, due to a rather small misalignment of the sample.
Because such a peak was not observed, we had good reasons to believe
that the samples were oriented with the accuracy better than 1
degree.

Using the model parameters determined from the analysis of the
frequency and field dependences of  $\chi'$ and  $\chi''$, we
obtained a satisfactory description of the temperature dependences
as well (see Fig.\ref{fig:10}). The transformations of the magnetic
field dependences of the out-of-phase susceptibility at the fixed
frequency in the vicinity of the crossing points with temperature
(Figs.\ref{fig:5},\ref{fig:7}) or concentration (Fig.\ref{fig:6})
can be explained as a result of strong variation of relaxation rates
with the magnetic field at these points.  If the effective
relaxation rate exceeds the frequency of the ac field,  $\chi''$ has
a dip, and if the relaxation rate is less than the frequency,
$\chi''$ has a peak at the crossing point. The corresponding "peak
to dip" transformations can be easily recognized in
Figs.\ref{fig:5}-\ref{fig:7}.

\section{Conclusion}

We conclude that the microscopic model of a linear non-resonant
response of the electron-nuclear subsystem in the highly diluted
paramagnetic crystal LiYF$_4$:Ho$^{3+}$ on the weak ac field has
been derived. The model operates with a few phenomenological
parameters introduced to account for a finite heat capacity of the
resonant phonons and cross-relaxation processes. The independent on
temperature finite phonon lifetimes of about 1 $\mu$s, comparable
with the phonon time of flight between the sample boundaries, have
been determined from fitting the results of calculations to the
measured frequency dependences of out-of-phase susceptibilities in
the samples with different holmium concentrations. The parameters of
the interaction between paramagnetic ions, which determine
cross-relaxation rates, were obtained from fitting the simulated
dependences of in-phase and out-of-phase susceptibilities on the
external magnetic field. The obtained value of the parameter
k$_{44}$, which defines co-tunneling processes at half-integer
crossing points in the spectrum of a pair of Ho$^{3+}$ ions, agrees
with the estimation based on the dipole-dipole mechanism of
inter-ion coupling. However, to present definite conclusions about
the most important terms in the Hamiltonian of interaction between
the holmium ions, which are responsible for the cross-relaxation
processes at integer crossing and anti-crossing points, it is
necessary to derive a more elaborated theory of cross-relaxation
than the high temperature approach of Ref.[5].

The model containing single-ion crystal-field including random
strains, electron-phonon transitions including bottleneck, hyperfine
and cross-spin interactions, is comprehensive and can be used to
predict the detailed magnetization dynamics in LiYF$_4$:Ho$^{3+}$
crystals at ultra low temperatures and at higher frequencies. Note
that half-integer transitions were observed at elevated temperatures
(above 1.5 K) only in ac susceptibility experiments or at fast
sweeping field in the low temperature (below 0.1 K) magnetization
measurements where the bottleneck effects stabilized an effective
spin-phonon temperature of half a Kelvin
[\onlinecite{Giraud2001,Giraud2003,barbara2004}]. In the present
theoretical approach half-integer transitions result from resonant
cross-relaxation transitions involving sublevels of the lower
singlet with the activation energy of $\sim$10 K, which is
satisfactory to fit the experimental data taken at liquid helium
temperatures. A more detailed analysis of the magnetization dynamics
in the sweeping fields will be presented in a separate paper. The
theory may be additionally tested by measurements of the ac
susceptibility and NMR in a tilted magnetic field. Finally, this
theoretical approach may be also expanded to other magnetic systems,
in particular, the single molecule magnets.

\section*{acknowledgements}
This work was supported by INTAS (project 03-51-4943) and the
Ministry of Education and Science of Russian Federation (project RNP
2.1.1.7348).

\end{document}